\documentclass[twocolumn,twoside]{revtex4}
\usepackage{graphicx}
\usepackage{fancyhdr}
\begin{document}

\title{Hadronic $D_{(s)}$ decays at BESIII}

%

\author{Z.~H.~Lu\\
on behalf of the BESIII Collaboration}
\affiliation{Institute of High Energy Physics, Beijing, People’s Republic of China, 100049}

\begin{abstract}
Charm hadronic decays are an important tool for understanding the the hadron spectroscopy and dynamics of the strong interaction in the low energy regime. BESIII experiment has collected a huge amount of data at various energy points around $D\bar{D}$ and $D_sD_s^*$ threshold. This report summarizes recent amplitude analyses and branching fraction measurements results of BESIII related to $D_{(s)}$ decays.
\end{abstract}

\maketitle

\thispagestyle{fancy}


\section{Introduction}
The ground states of charmed hadrons $e.g.$, $D_{(s)}$, can only decay weakly and so precision studies of these charm decays provide important constraints on the weak interaction~\cite{Fotopoulos:2008ka}. Furthermore, as the strong force is always involved in the decays of the charmed hadrons and the formation of the final-state hadrons, precise measurements of the decay properties allow tests of non-perturbative quantum chromodynamics (QCD) calculations. To investigate the possible intermediate states in $D_{(s)}$ decays, amplitude analysis is a powerful tool to help understand the different decay mechanisms, such as weak-annihilation, W-exchange and final-state scattering.

The BESIII detector is a magnetic spectrometer~\cite{BESIII:2009fln,BESIII:2020nme} located at the Beijing Electron Positron Collider (BEPCII)~\cite{Yu:IPAC2016-TUYA01}, which produces charmed hadrons near their mass threshold; this allows exclusive reconstruction of their decay products with well-determined kinematics. Up to now, BESIII has collected data corresponding to the integrated luminosities of 2.93 ${\rm fb^{-1}}$ and 6.32 ${\rm fb^{-1}}$ at the center-of-mass energies ($\sqrt{s}$) of 3.773 and 4.178-4.226 GeV, respectively. Based on these data sets, many world-leading results have been published. In this paper, charge conjugate states are implied.

\section{Strategy}
The tag method~\cite{MARK-III:1985hbd} is employed to select clean signal samples of $e^+e^- \to \psi(3770) \to D^0\bar{D}^0$ and $e^+e^- \to D_s^{*\pm}D_s^\mp \to \gamma D_s^+D_s^-$ in the following analyses. In this method, a single-tag (ST) candidate requires a reconstructed $\bar{D}$($D_{s}^{-}$) decay to any of final states. A double-tag (DT) candidate requires that the $D$($D^+_s$) is reconstructed in the signal mode in addition to the $\bar{D}$($D_{s}^{-}$) decays to one of the tag modes.

In the amplitude analysis, the total amplitude $\mathcal{M}$ is treated with the isobar model, which uses the coherent sum of the amplitudes of the intermediate processes, $\mathcal M(p) = \sum{c_n\mathcal A_n(p)}$, where $c_n = \rho_ne^{i\phi_n}$ is the corresponding complex coefficient. The magnitude $\rho_n$ and phase $\phi_n$ are the free parameters to be determined in the fit. The amplitude of the $n^{\rm th}$ intermediate state ($\mathcal A_n$) is
\begin{equation}
	\mathcal A_n(p) = P_n(p)S_n(p)F^r_n(p)F^D_n(p).
\end{equation}
Here, $P_n(p)$ is the propagator of the intermediate resonance, $S_n(p)$ is the spin factor~\cite{covariant-tensors}, $F^{r}_n(p)$ and $F^{D}_n(p)$ are the Blatt-Weisskopf barrier factors~\cite{PhysRevD.104.012016} for the intermediate resonance and $D_{(s)}$, respectively.

\section{Amplitude analysis}

\subsection{$D_s^+ \to K_S^0K_S^0\pi^+$}
\begin{figure}[h]
\centering
\includegraphics[width=70mm]{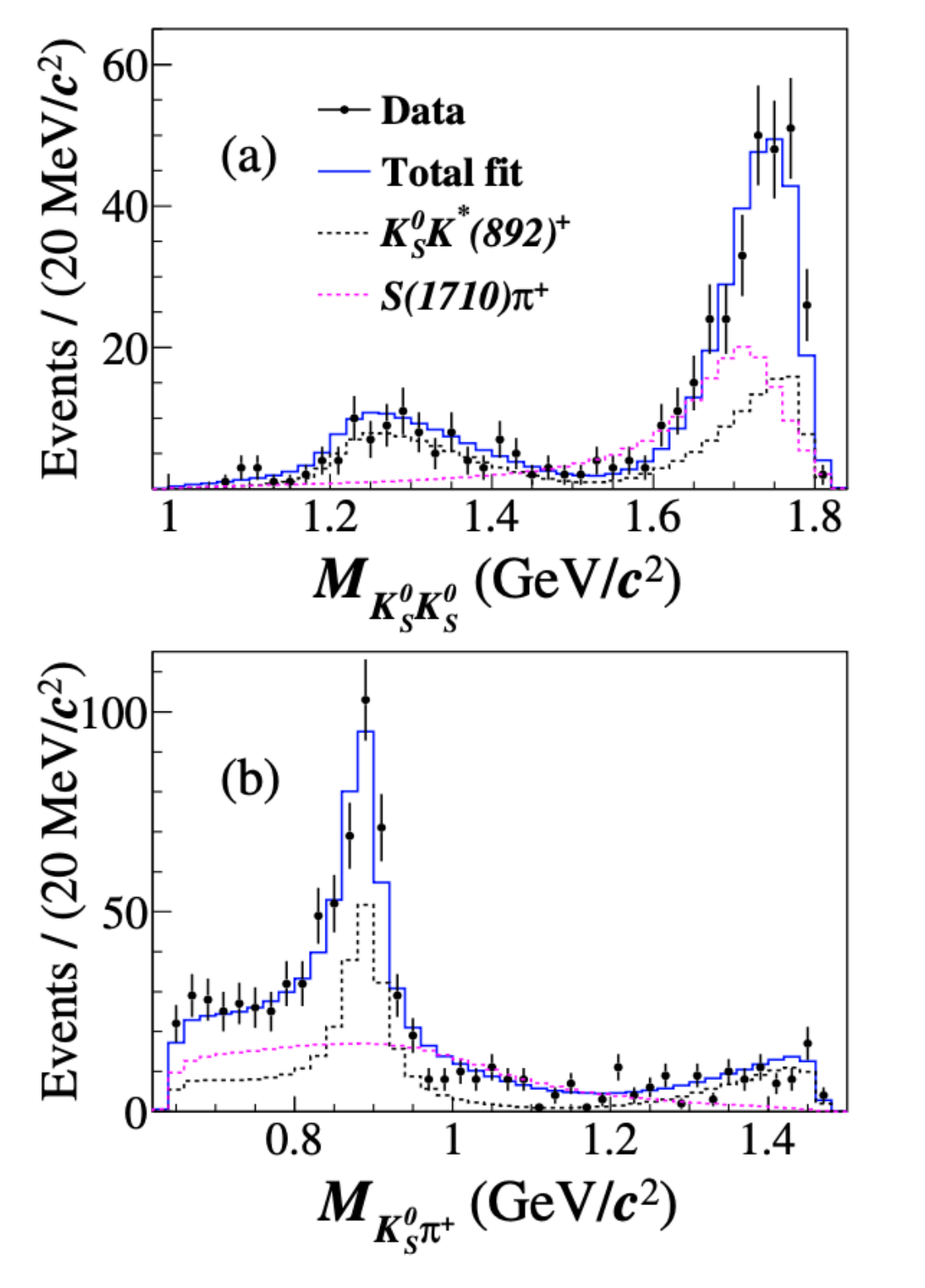}
\caption{Distribution of (a) $M_{K_S^0K_S^0}$ and (b) $M_{K_S^0\pi^+}$ from the nominal fit. The distribution of $M_{K_S^0\pi^+}$ contains two entries per event, one for each $K_S^0$. The data samples are represented by points with uncertainties and the fit results by the blue lines. Colored dashed lines show the individual components of the fit model. Due to interference effects, the fit results are not necessarily equal to the sum of the components.} \label{KSKSpi}
\end{figure}

The $\textstyle\mathit{BABAR}$ collaboration recently claimed the observation of a new $a_0(1700)^\pm$ resonance with a mass approximately 1.7 ${\rm GeV/}c^2$~\cite{BaBar:2021fkz}, one can also expect this new resonance exists in the $KK$ spectrum. An excellent way to search for $a_0(1710)$, which is suppose to be an isospin one partner of the $f_0(1710)$, is the simultaneous fit of the decays involving $K^+K^-$ and $K_S^0K_S^0$, e.g., the combination measurement of $D_s^+ \to K^+K^-\pi^+$~\cite{BESIII:2020ctr} and $D_{s}^{+} \to K_{S}^{0}K_{S}^{0}\pi^{+}$. Hence, the BESIII collaboration performed an amplitude analysis of the decay $D_{s}^{+} \to K_{S}^{0}K_{S}^{0}\pi^{+}$ for the first time, using 412 DT events with a signal purity of 97.3\%~\cite{BESIII:2021anf}. Figure~\ref{KSKSpi} shows the projections of the nominal fit result. Due to destructive interference between $a_0(980)^0$ and $f_0(980)$ in decays to two neutral kaons, almost no signal populates the region below 1.1~GeV/$c^2$ in the $K_{S}^0K_{S}^0$ mass spectrum. The same interference term would then be constructive, implying the existence of $a_0(1710)^0$. From the analysis, the branching fraction (BF) of $D_s^+ \to S(1710)\pi^+$ [where $S(1710)$ denotes an admixture of $a_0(1710)$ and $f_0(1710)$] is determined to be $(3.1 \pm0.3_{\rm stat.} \pm 0.1_{\rm syst.})\times 10^{-3}$, which is one order of magnitude larger than the expectation. Moreover, the absolute BF of $D_s^+ \to K_S^0K_S^0\pi^+$ is also measured, which is $(0.68\pm0.04_{\rm stat.}\pm0.01_{\rm syst.})\%$.

\subsection{$D_s^+ \to K_S^0K^+\pi^0$}
\begin{figure*}[t]
\centering
\includegraphics[width=170mm]{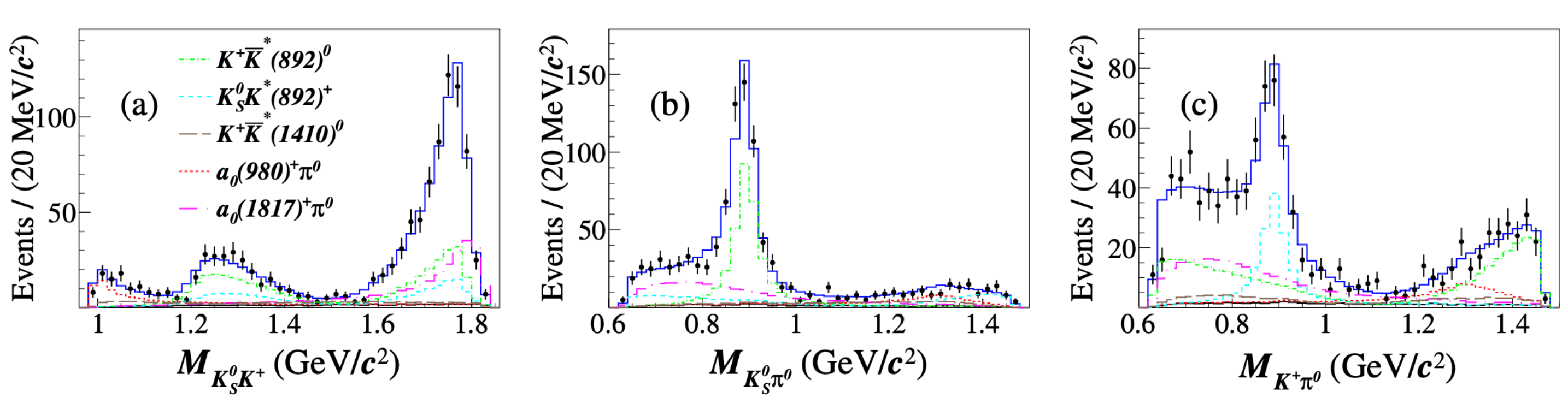}
\caption{The projections of the Dalitz plot onto (a) $M_{K_S^0K^+}$, (b) $M_{K_S^0\pi^0}$, and (c) $M_{K^+\pi^0}$. The data samples are represented by points with error bars, the fit results by blue lines, and backgrounds by black lines. Colored dashed lines show the components of the fit model. Due to interference effects, the fit results are not necessarily equal to the sum of the components.} \label{KSKpi0}
\end{figure*}

We also performed a amplitude analysis of the decay $D_s^+ \to K_S^0K^+\pi^0$~\cite{BESIII:2022wkv}, which provides an ideal opportunity to pin down the nature of the new $a_0$(s) (called $a_0(1817)$ in its Letter). The analysis was done with 1050 DT events with a signal purity of 94.7\%, and the projections of the fit result are shown at Fig.~\ref{KSKpi0}. In this analysis, the statistical significance of $D_{s}^{+} \to a_0(1817)^+\pi^{0}$ is found to be greater than $10\sigma$. The mass and width of new $a_0$ are measured to be ($1.817 \pm 0.008_{\rm stat.} \pm 0.020_{\rm syst.}$)~GeV/$c^2$ and ($0.097 \pm 0.022_{\rm stat.} \pm 0.015_{\rm syst.}$)~GeV/$c^{2}$, respectively. The BF of $D_{s}^{+} \to a_0(1817)^+\pi^{0}$ with $a_0(1817)^+\to K_S^0K^+$ is $(3.44 \pm 0.52_{\rm stat.} \pm 0.32_{\rm syst.})\times 10^{-3}$. The results of amplitude analysis of $D_s^+ \to K_S^0K_S^0\pi^+$ and $D_s^+ \to K_S^0K^+\pi^0$ confirm the existence of a new $a_0$ triple. The measured BF of $D_{s}^{+} \to a_0(1817)^+\pi^{0}$ is roughly consistent with the prediction~\cite{Dai:2021owu} assuming $a_0(1817)^+$ is the candidate of isovector partner of $f_0(1710)$, but the mass is about 100~MeV/$c^2$ greater than what it is supposed to be. This higher mass may imply $a_0(1817)$ is the isovector partner of the $X(1812)$ instead~\cite{Guo:2022xqu}. In addition, the BF of $D_s^+ \to K_S^0K^+\pi^0$ is determined to be $(1.46\pm0.06_{\rm stat.}\pm0.05_{\rm syst.})\%$, and the ratio $\frac{\mathcal{B}(D_{s}^{+} \to \bar{K}^{*}(892)^{0}K^{+})}{\mathcal{B}(D_{s}^{+} \to \bar{K}^{0}K^{*}(892)^{+})}$ is also calculated, which is $2.35^{+0.42}_{-0.23\text{stat.}}\pm 0.10_{\rm syst.}$.

\subsection{$D_s^+ \to \pi^+\pi^0\pi^0$}
The BF of $D_s^+ \to f_{0(2)}\pi^+$ reported from the amplitude analysis of $D_s^+\to\pi^+\pi^+\pi^-$ decays has large discrepancies~\cite{ParticleDataGroup:2020ssz,E687:1997jvh,E791:2000lzz} with that measured from the $D_s^+\to K^+K^-\pi^+$ decays, in which the $f_{0(2)}$ contributions may suffer from the contaminations of $a_0(980) \to K^+K^-$ or $\rho \to \pi^+\pi^-$. Therefore, the $D_s^+ \to \pi^+\pi^0\pi^0$ decays offer a cleaner environment due to absence of these contributions. The BESIII collaboration recently published the first amplitude analysis of $D^+_s\to \pi^+\pi^0\pi^0$ and a more precise measurement of its absolute BF using 572 DT events with a signal purity of about 78\%~\cite{BESIII:2021eru}. From the analysis, the BF of $D_{s}^{+} \to f_0(980)\pi^{+}, f_0(980)\to\pi^{0}\pi^{0}$ is measured for the first time, which is $(0.28\pm 0.04_{\text{stat.}}\pm 0.04_{\text{syst.}})\%$. Furthermore, the ratio of $\frac{f_{0(2)}\to\pi^+\pi^-}{f_{0(2)}\to\pi^0\pi^0}$ is examined along with the BFs of $f_{0(2)}\to\pi^+\pi^-$ and $f_{0(2)}\to\pi^0\pi^0$~\cite{ParticleDataGroup:2020ssz}. The ratios for $f_0(980)$, $f_0(1370)$, and $f_2(1270)$ are $2.2\pm 0.5$, $2.7\pm 1.4$, and $2.4\pm 1.8$, respectively, which are consistent with the expectation value of 2 based on isospin symmetry. Our results are consistent with those from $D^+_s\to \pi^+\pi^+\pi^-$. In addition, no significant signal of $f_0(500)$ is observed. The BF of $D_{s}^{+} \to \pi^{+}\pi^{0}\pi^{0}$ is also determined to be $(0.50\pm 0.04_{\text{stat.}}\pm 0.02_{\text{syst.}})\%$, which is improved by about a factor of two compared to the PDG value~\cite{ParticleDataGroup:2020ssz}.

\subsection{$D_s^+ \to \pi^+\pi^0\eta^\prime$}
Previously, BESIII reported the BF measurement of $D_s^+ \to \rho^+\eta^\prime$~\cite{BESIII:2015rrp}, which is larger than the theoretical predictions by around 2$\sigma$~\cite{Fu-Sheng:2011fji,Qin:2013tje}. Using 411 DT events with a purity of 96.1\%, we performed the first amplitude analysis of $D_s^+ \to \pi^+\pi^0\eta^\prime$~\cite{BESIII:2022ewq}. The analysis shows that the dominant intermediate process is $D^+_s \to \rho^+ \eta^\prime$ and the significances of other resonant and nonresonant processes are all less than 3$\sigma$. The upper limits on the BFs of S-wave and P-wave nonresonant components are set to 0.10\% and 0.74\% at the 90\% confidence level, respectively. In addition, the BF of the $D^+_s\to \pi^+\pi^0\eta^\prime$ decay is measured to be $(6.15 \pm 0.25_{\text{stat.}} \pm 0.18_{\text{syst.}})\%$, which receives significant contribution only from $D^+_s \to \rho^+ \eta^\prime$ according to the amplitude analysis. This result is more than 3$\sigma$ above current theoretical predictions and suggests that other contributions, such as, QCD flavor-singlet hairpin amplitude, should be taken into account.

\subsection{$D_s^+ \to K^+\pi^+\pi^-$}
\begin{figure}[h]
\centering
\includegraphics[width=85mm]{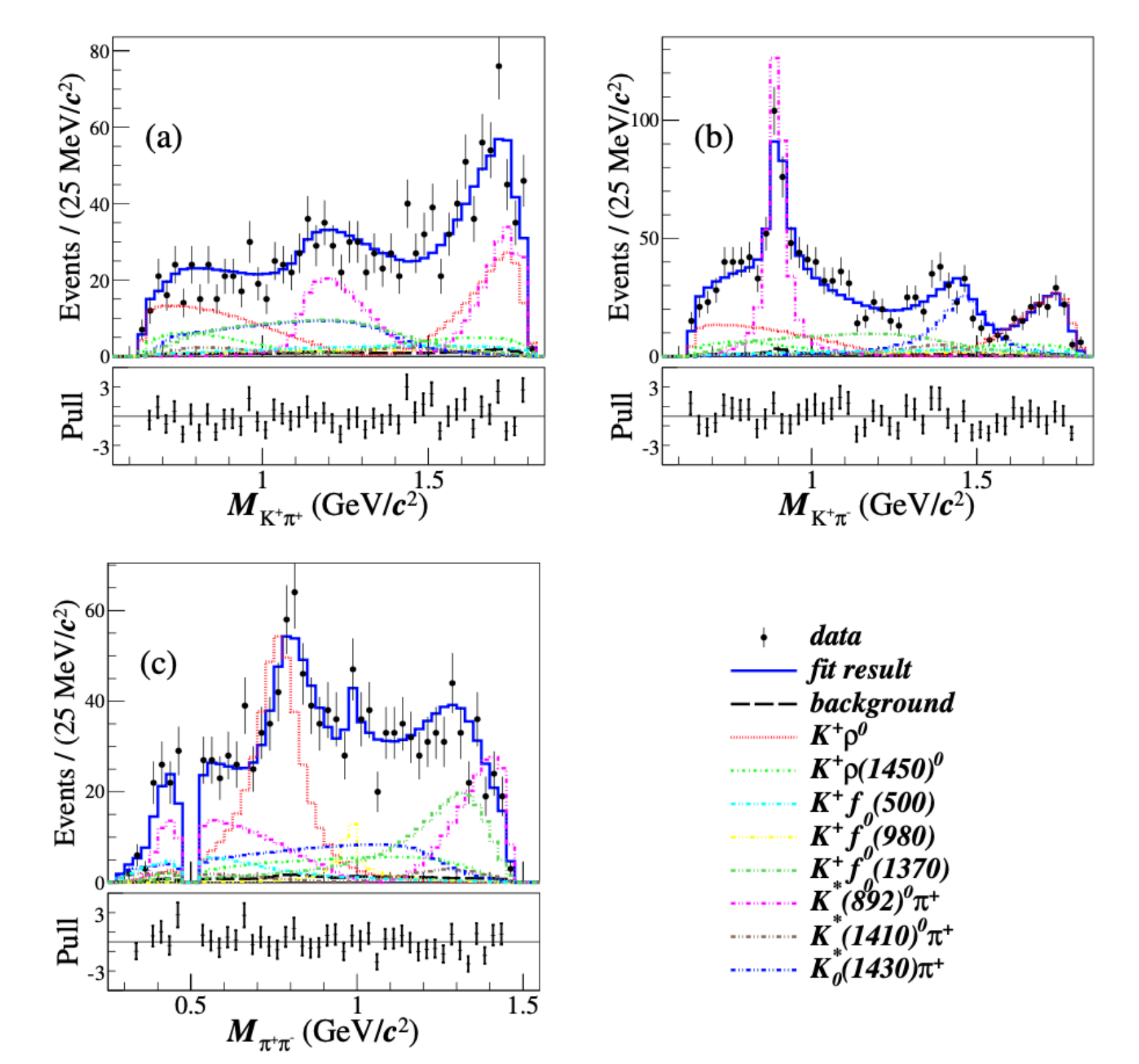}
\caption{The projections on (a) $M_{K^+\pi^+}$, (b) $M_{K^+\pi^-}$, and (c) $M_{\pi^+\pi^-}$ of the nominal fit. The data samples are represented by points with error bars, the fit results by the solid blue lines, and the background estimated from the inclusive MC samples by the black dashed lines. Colored curves show the components of the fit model. Due to interference effects, the total is not necessarily equal to the linear sum of the components. Pull projections are shown beneath each distribution.} \label{Kpipi}
\end{figure}

An amplitude analysis of $D_s^+ \to K^+\pi^+\pi^-$ has been performed by the FOCUS collaboration with 567 events~\cite{FOCUS:2004muk}, the data samples they used suffered from high contaminations and it is difficult to search for potential intermediate states. Using 1356 DT events with a signal purity of 95\%, the BESIII collaboration reported the updated amplitude analysis result of this channel~\cite{BESIII:2022vaf}. Figure~\ref{Kpipi} shows the projections of the nominal fit. The dominant states are $D_s^+ \to K^+\rho^0$ and $D_s^+ \to K^*(892)^0\pi^+$, whose BFs are determined to be $(1.96 \pm 0.19_{\text{stat.}} \pm 0.23_{\text{syst.}})\times 10^{-3}$ and $(1.85 \pm 0.12_{\text{stat.}} \pm 0.13_{\text{syst.}})\times 10^{-3}$, respectively. The contribution from nonresonant are removed from the FOCUS model and the $f_0(500)$, $f_0(980)$, and $f_0(1370)$ are observed in this channel for the first time. Moreover, the BF of $D_s^+ \to K^+\pi^+\pi^-$ is also measured to be $(6.11\pm0.18_{\rm stat.}\pm0.11_{\rm syst.})\times 10^{-3}$, which is improved by about a factor of 2 compared to the world average value~\cite{ParticleDataGroup:2020ssz}. In addition, the asymmetry of the BFs of $D_s^+ \to K^+\pi^+\pi^-$ and $D_s^- \to K^-\pi^-\pi^+$ is determined to be $(3.3\pm{{3.0}}_{\rm stat.}\pm1.3_{\rm syst.})\%$. No indication of $\mathit{{CP}}$ violation is found. 

\section{branching fraction measurement}
\subsection{$D^0 \to \omega\phi$}
To date, all experimental measurements of the helicity in $D^0 \to VV$ decays have been contrary to theoretical predictions, and the puzzle needs to be confirmed with more precise measurements and validated using more decay modes. Unlike the decay modes such as $D^0 \to \rho^0\rho^0$ which require complicated and model dependent analyses of multibody decays, the narrow widths of $\omega$ and $\phi$ signals allow for a straightforward signal extraction in $D^0 \to \omega\phi$. The BESIII collaboration published the first measurement of polarizations in the decay $D^0 \to \omega\phi$~\cite{BESIII:2021raf}, in which the $\phi$ and $\omega$ mesons are found to be transversely polarized, which is the same as that observed in the decay $D^0 \to \bar{K}^{*0}\rho^0$ but contradicts predictions from the naive factorization~\cite{Cheng:2010rv} and Lorentz invariant-based symmetry~\cite{Hiller:2013cza} models. Furthermore, the longitudinal polarization fraction is set to be less than 0.24 on the 95\% confidence level. In addition, the BF of $D^0 \to \omega\phi$ is measured to be $(6.48\pm 0.96_{\rm stat.} \pm 0.40_{\rm syst.})\times10^{-3}$ with a significance of 6.3$\sigma$, which is consistent with predictions from Refs.~\cite{Hiller:2013cza,Kamal:1990ky} but inconsistent with Refs.~\cite{Kamal:1990ky,Jiang:2017zwr}.

\section{Summary}
Based on $e^+e^-$ annihilation data corresponding to a total integrated luminosity of 2.93 ${\rm fb^{-1}}$ and 6.32 ${\rm fb^{-1}}$ at $\sqrt{s}$ = 3.773 and 4.178-4.226 GeV with the BESIII detector, we have made many great achievements of the $D_{(s)}$ decays. We established the isospin-one particle $a_0(1710)$ with the combination measurements of $D_s^+ \to K_S^0K_S^0\pi^+$ and $D_s^+ \to K_S^0K^+\pi^0$, and the amplitude analyses of $D_s^+ \to \pi^+\pi^0\pi^0$, $D_s^+ \to \pi^+\pi^0\eta^\prime$ and $D_s^+ \to K^+\pi^+\pi^-$ can help test various theories. Furthermore, the large data sets make it possible to perform first measurement of $D^0 \to \omega\phi$, and provide important information to understand of the underlying dynamics in charmed meson decays and also may help in searches for new physics. In the future, more analyses will be conducted with a total integrated luminosity of 20 ${\rm fb^{-1}}$ at $\sqrt{s}$ = 3.773 GeV~\cite{BESIII:2020nme}.


\begin{acknowledgments}
We would like to thanks for the strong support from the staff of BEPCII and the IHEP computing center, and also thank our BESIII collaborators for contributing to this proceeding. At last, thank FPCP2022 organizing committee for holding this wonderful conference.

\end{acknowledgments}

\bigskip 

\begin{thebibliography}{99} 

\bibitem{Fotopoulos:2008ka} A.~Fotopoulos and M.~Tsulaia, Int. J. Mod. Phys. A {\bf 24}, 1 (2009).

\bibitem{BESIII:2009fln} M.~Ablikim {\em et al.} (BESIII Collaboration), Nucl. Instrum. Meth. A {\bf 614}, 345 (2010).

\bibitem{BESIII:2020nme} M.~Ablikim {\em et al.} (BESIII Collaboration), Chin. Phys. C {\bf 44}, 040001 (2020).

\bibitem{Yu:IPAC2016-TUYA01} C.~Yu {\em et al.} Proceedings of IPAC2016, Busan, Korea, 2016, doi:10.18429/JACoW-IPAC2016-TUYA01.

\bibitem{MARK-III:1985hbd} R.~M.~Baltrusaitis {\em et al.} (MARK-III Collaboration), Phys. Rev. Lett.
{\bf 56}, 2140 (1986).

\bibitem{covariant-tensors} B. S. Zou and D. V. Bugg, Eur. Phys. J. A {\bf 16}, 537 (2003).

\bibitem{PhysRevD.104.012016} M.~Ablikim {\em et al.} (BESIII Collaboration), Phys. Rev. D {\bf 104}, 012016
(2021).

\bibitem{BaBar:2021fkz} J.~P.~Lees {\em et al.} (BaBar Collaboration), Phys. Rev. D {\bf 104}, 072002 (2021).

\bibitem{BESIII:2020ctr} M.~Ablikim {\em et al.} (BESIII Collaboration), Phys. Rev. D {\bf 104}, 012016 (2021).

\bibitem{BESIII:2021anf} M.~Ablikim {\em et al.} (BESIII Collaboration), Phys. Rev. D {\bf 105}, L051103 (2022).

\bibitem{BESIII:2022wkv} M.~Ablikim {\em et al.} (BESIII Collaboration), arxiv: 2110.07650.

\bibitem{Dai:2021owu} L.~R.~Dai, E.~Oset, and L.~S.~Geng, Eur. Phys. J. C {\bf 82}, 225 (2022).

\bibitem{Guo:2022xqu} D.~Guo, W.~Chen, H.~X.~Chen, X.~Liu, and S.~L.~Zhu, Phys. Rev. D {\bf 105}, 114014 (2022).

\bibitem{ParticleDataGroup:2020ssz} P.~A.~Zyla {\em et al.} (Particle Data Group), PTEP {\bf 2020}, 083C01 (2020).

\bibitem{E687:1997jvh} P.~L.~Frabetti {\em et al.} (E687 Collaboration), Phys. Lett. B {\bf 407}, 79 (1997).

\bibitem{E791:2000lzz} E.~M.~Aitala {\em et al.} (E791 Collaboration), Phys. Rev. Lett. {\bf 86}, 765 (2001).

\bibitem{BESIII:2021eru} M.~Ablikim {\em et al.} (BESIII Collaboration), JHEP {\bf 01}, 052 (2022).

\bibitem{BESIII:2015rrp} M.~Ablikim {\em et al.} (BESIII Collaboration), Phys. Lett. B {\bf 750}, 466 (2015).

\bibitem{Fu-Sheng:2011fji} Y.~F.~Sheng, X.~X.~Wang, and C.~D.~L\"u, Phys. Rev. D {\bf 84}, 074019 (2011).

\bibitem{Qin:2013tje} Q.~Qin, H.~N.~Li, C.~D.~L\"u, and F.~S.~Yu, Phys. Rev. D {\bf 89}, 054006 (2014).

\bibitem{BESIII:2022ewq} M.~Ablikim {\em et al.} (BESIII Collaboration), JHEP {\bf 04}, 058 (2022).

\bibitem{FOCUS:2004muk} J.~M.~Link {\em et al.} (FOCUS Collaboration), Phys. Lett. B {\bf 601}, 10 (2004).

\bibitem{BESIII:2022vaf} M.~Ablikim {\em et al.} (BESIII Collaboration), arxiv: 205.08844.

\bibitem{BESIII:2021raf} M.~Ablikim {\em et al.} (BESIII Collaboration), Phys. Rev. Lett. {\bf 128}, 011803 (2022).

\bibitem{Cheng:2010rv} H.~Y.~Cheng and C.~W.~Chiang, Phys. Rev. D {\bf 81}, 114020 (2010).

\bibitem{Hiller:2013cza} G.~Hiller and R.~Zwicky, JHEP {\bf 03}, 042 (2014).

\bibitem{Kamal:1990ky} A.~N.~Kamal, R.~C.~Verma, and N.~Sinha, Phys. Rev. D {\bf 43}, 843 (1991).

\bibitem{Jiang:2017zwr} H.~Y.~Jiang, F.~S.~Yu, Q.~Qin, H.~N.~Li, and C.~D.~L\"u, Chin. Phys. C {\bf 42}, 063101 (2018).


\end{thebibliography}


\end{document}